\documentclass[superscriptaddress,twocolumn,aps,prl,floatfix,showpacs]{revtex4}
\usepackage{graphicx}
\usepackage{amsmath,amssymb,amsfonts}
\usepackage{dcolumn}% Align table columns on decimal point
\usepackage{bm}% bold math

\begin{document}
\title{Skyrmions in a Doped Antiferromagnet}
\author{I. Rai\v{c}evi\'{c}}
\affiliation{National High Magnetic Field Laboratory, Florida State University, Tallahassee, FL 32310, USA}
\affiliation{Department of Physics, Florida State University, Tallahassee, FL 32306, USA}
\author{Dragana Popovi\'c}
\affiliation{National High Magnetic Field Laboratory, Florida State University, Tallahassee, FL 32310, USA}
\affiliation{Department of Physics, Florida State University, Tallahassee, FL 32306, USA}
\author{C. Panagopoulos}
\affiliation{Department of Physics, University of Crete and FORTH,
71003 Heraklion, Greece} \affiliation{Division of Physics and
Applied Physics, Nanyang Technological University, Singapore}
\author{L.~Benfatto}
\affiliation{CNR-ISC and Department of Physics, ``Sapienza'' University of
Rome, 00185 Rome, Italy}
\author{M. B. Silva Neto}
\affiliation{Instituto de F\' \i sica,
Universidade Federal do Rio de Janeiro, CEP 21945-972, Rio de
Janeiro - RJ, Brasil}
\author{E. S. Choi}
\affiliation{National High Magnetic Field Laboratory, Florida
State University, Tallahassee, FL 32310, USA}
\author{T. Sasagawa}
\affiliation{Materials and Structures Laboratory, Tokyo Institute
of Technology, Kanagawa 226-8503, Japan}

\begin{abstract}
 Magnetization and magnetoresistance have been measured in insulating
 antiferromagnetic La$_{2}$Cu$_{0.97}$Li$_{0.03}$O$_{4}$ over a wide range
 of temperatures, magnetic fields, and field orientations.  The
 magnetoresistance step associated with a weak ferromagnetic transition
 exhibits a striking nonmonotonic temperature dependence, consistent with
 the presence of skyrmions.

\end{abstract}
% PACS options: 75.47.-m 	Magnetotransport phenomena; materials for magnetotransport
%               75.47.Lx 	Magnetic oxides
%               75.50.-y 	Studies of specific magnetic materials
%               75.50.Ee 	Antiferromagnetics
%               72. 	    Electronic transport in condensed matter
%               72.20.My 	Galvanomagnetic and other magnetotransport effects
%               72.80.-r 	Conductivity of specific materials
%               72.80.Ga 	Transition-metal compounds
%

\maketitle

A remarkable manifestation of complexity in magnetic systems is the emergence of topologically non-trivial arrangements of spins, such as skyrmions \cite{physicstoday}.  These are ``knots'' in an otherwise ordered spin texture, which behave as excitations with particlelike properties.  Skyrmions are stabilized by a magnetic field $B$ in several ferromagnetic (FM) metals \cite{2DEG1,2DEG2,manganites,pfleiderer09}, where they manifest themselves in the electronic transport \cite{2DEG1,2DEG2,manganites,pfleiderer_prl09,ong_prl09}, or they form a detectable periodic skyrmion lattice \cite{pfleiderer09}.  Skyrmions have been predicted to emerge also in the ground state of doped
antiferromagnetic (AFM) insulators \cite{shraiman90,gooding91,haas96}, but the identification of such isolated skyrmions is a challenge.  Neutron scattering, for example, would not be a definitive probe, since skyrmions here do not form a lattice, while their possible signatures on transport may be screened by the insulating character of the carriers.

In a FM system, a charge carrier with the spin aligned to the magnetic background preserves its metallic character with a mass renormalization due to scattering by low-energy spin waves. Thus, the more complex spin excitations associated with topologically nontrivial magnetic textures have characteristic signatures in transport.  This is indeed the case for skyrmions in the polarized quantum Hall state of a two-dimensional electron gas
\cite{2DEG1,2DEG2}, in colossal magnetoresistance (MR) manganites \cite{manganites}, and in the three-dimensional FM MnSi \cite{pfleiderer_prl09,ong_prl09}.  In a doped AFM insulator, the description of transport is complex already in the topologically trivial AFM ground state, since the carriers cannot move without inducing spin-flip scattering \cite{kane89,shraiman90}.  If, however, the external $B$ field causes a change in transport that depends
on the configuration of the spin background, the MR becomes a key probe to identify signatures of anomalous spin textures. This situation can be realized in the AFM insulator La$_{2}$CuO$_{4}$, lightly hole doped with Li.  Here
we show that the striking nonmonotonic temperature ($T$) dependence of the MR step associated with a $B$-induced
magnetic transition is consistent with the skyrmion formation.

\begin{figure}

\begin{minipage}[c]{0.2\textwidth}
\hspace*{-0.26in}\includegraphics[width=4.25cm]{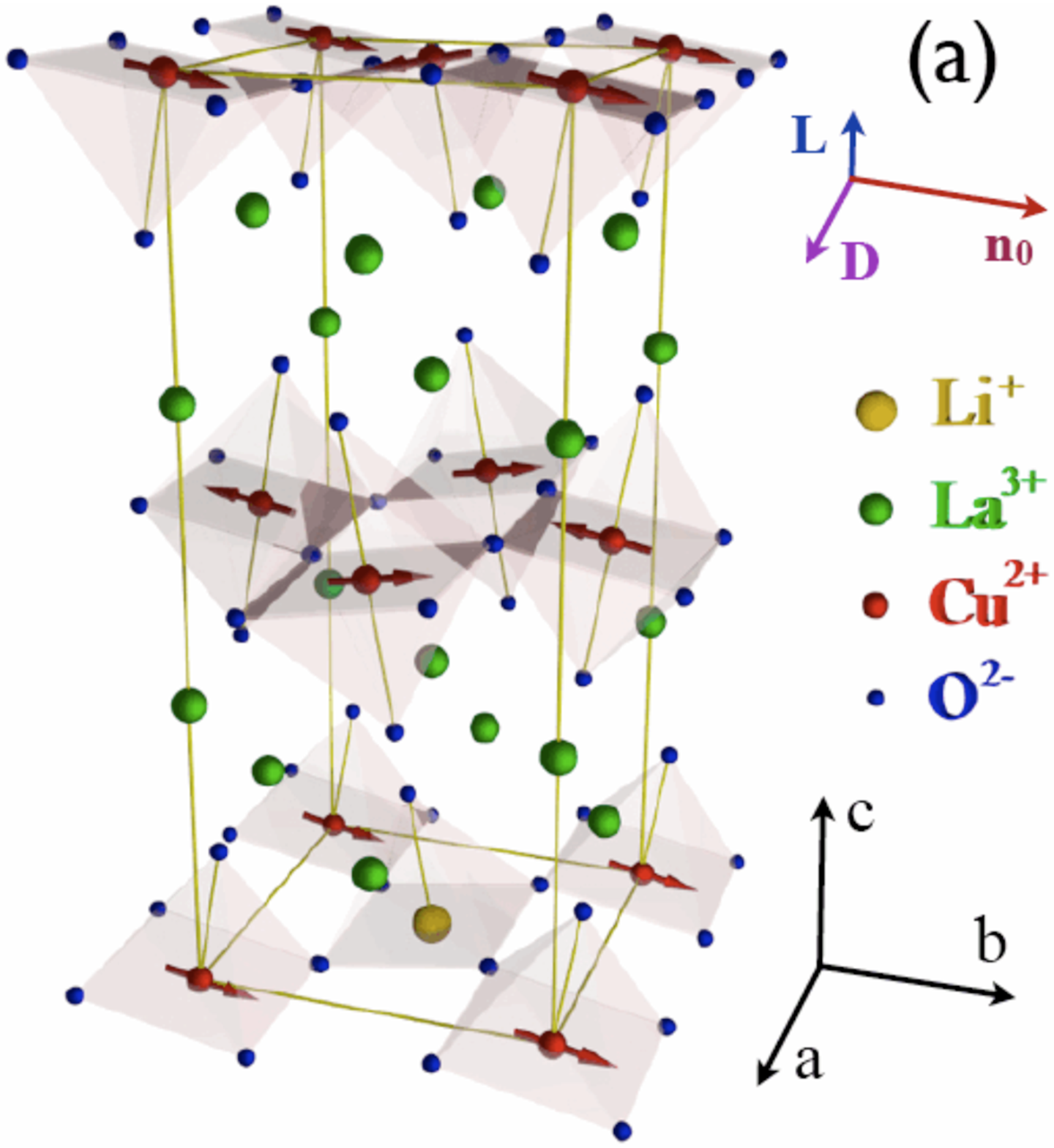}
\end{minipage}
\begin{minipage}[c]{0.2\textwidth}
\includegraphics[width=3.3cm]{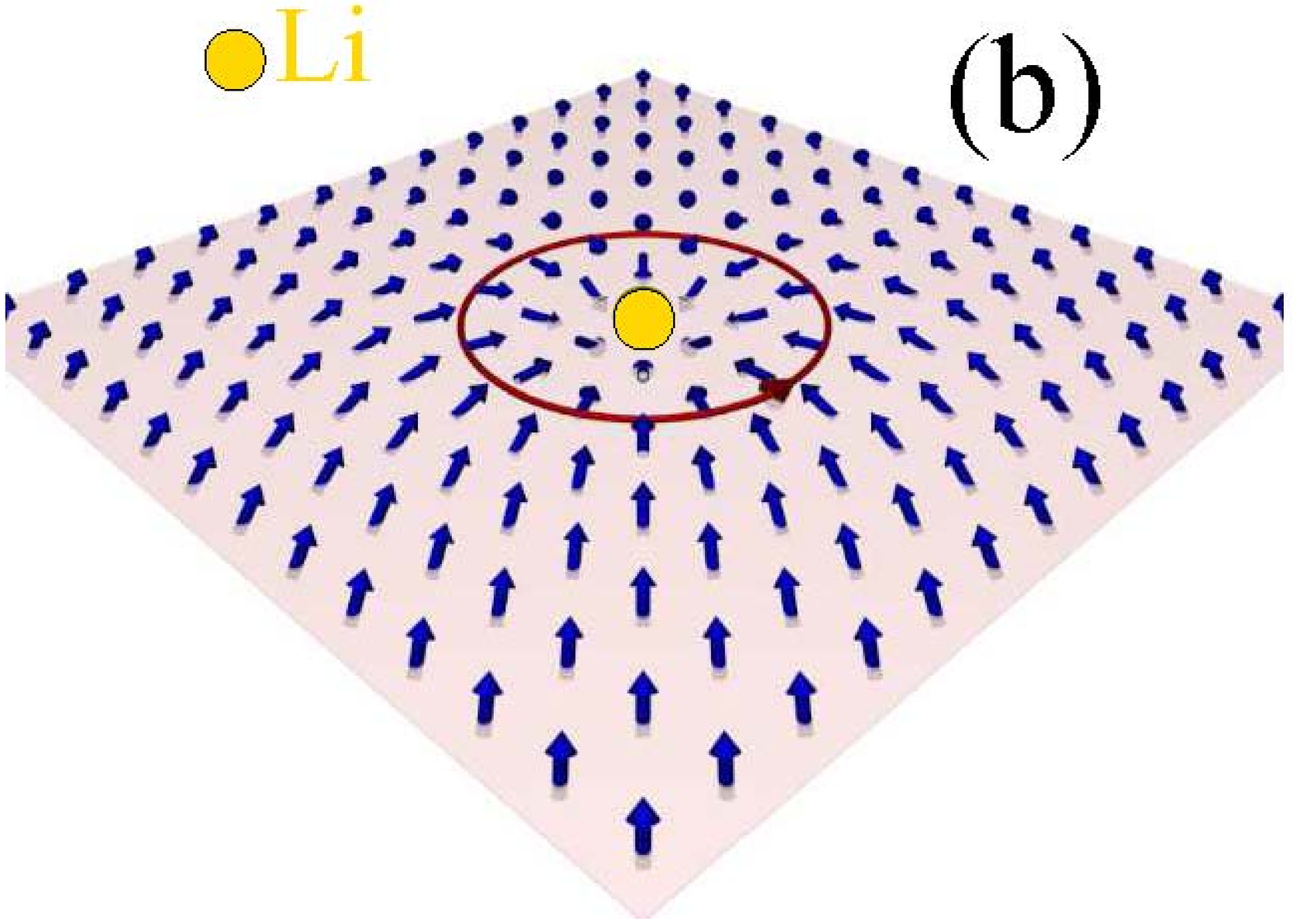}\\
\hspace*{0in}\includegraphics[width=3.5cm]{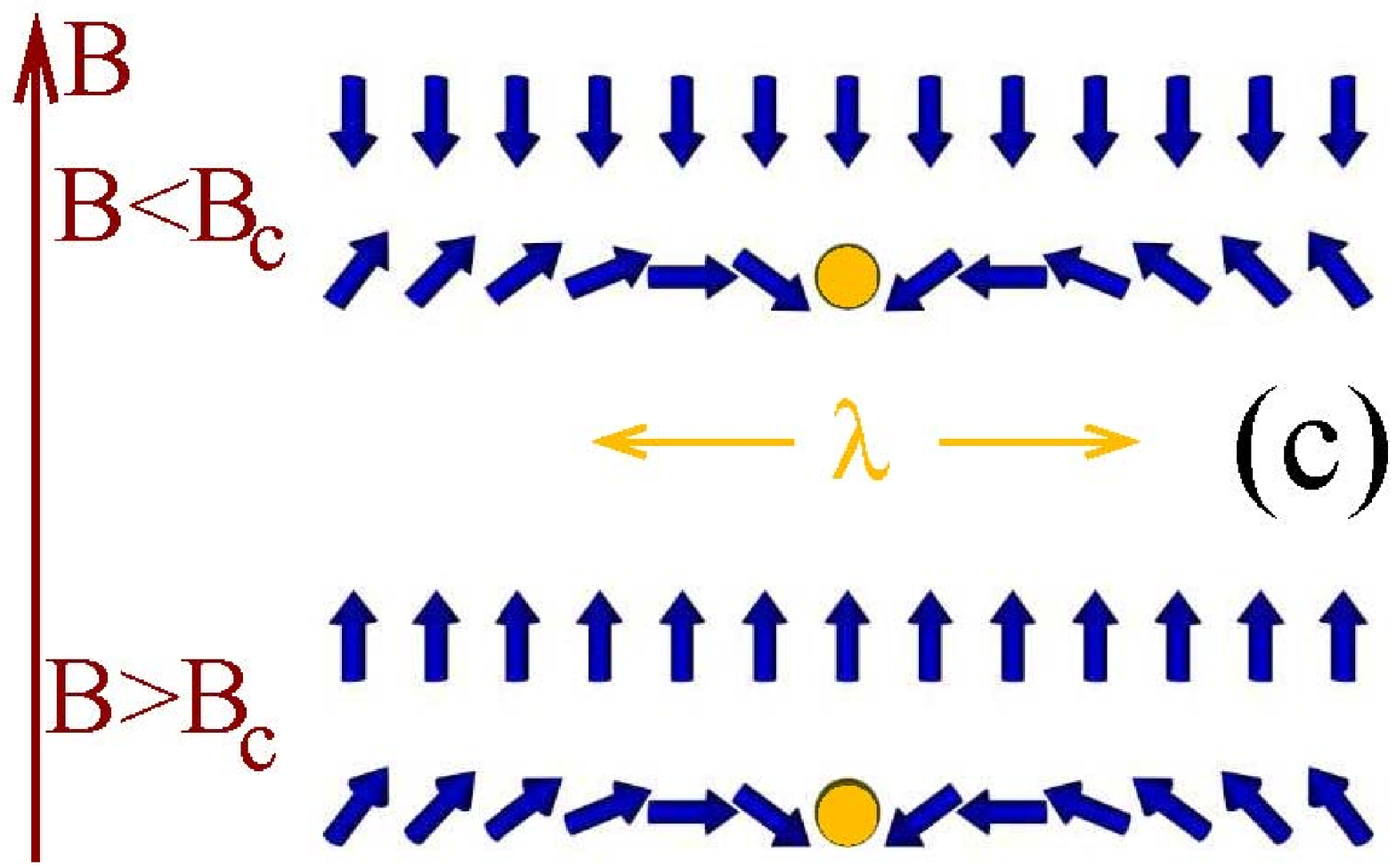}
\end{minipage}

\caption {(color online)
(a) Orthorhombic crystal structure of La$_{2}$CuO$_{4}$; Li replaces Cu in CuO$_{2}$ ($ab$) planes. The spins
on the Cu atoms have the AFM order with the large staggered component ${\bf n}_0$ along $b$ and small FM components ${\bf L}={\bf D}\times {\bf n}_0$ along $c$  (${\bf D}$, DM vector, aligned along $a$).  The weak FM moments have staggered order along $c$ at $B=0$.  (b) This AFM background becomes deformed near Li sites, via skyrmion formation.  For simplicity, we show the skyrmion configuration of the FM moments only.  The circle represents the skyrmion core size $\lambda$.  (c) Weak FM moments around the Li site across the weak FM transition: due to skyrmion formation, spins have a FM order inside a distance of the order of $\lambda$ at $B<B_c$ and AFM outside it, while the reverse is true for $B>B_c$.}
\label{fig:CrystalStructure}
\end{figure}

La$_{2}$CuO$_{4}$, the parent material of La$_{2}$Cu$_{1-x}$Li$_x$O$_{4}$, is a Mott insulator with a residual AFM coupling between the S = 1/2 spins located at the Cu$^{2+}$ ions, forming a nearly square lattice with a small
orthorhombic distortion \cite{kastner98} [Fig.~\ref{fig:CrystalStructure}(a)].
%
%%%%%%%%%%%%%%%%%%%%%%%%%%%%%%%%%%%%%%%%%%%%%%%%%%%%%%%%%%%%%%%%%%FIG1
%%%%%%%%%%%%%%%%%%%%%%%%%%%%%%%%%%%%%%%%%%%%%%%%%%%%%%%%%%%%%%%%%%FIG
%
A small antisymmetric Dzyaloshinskii-Moriya (DM) exchange between neighboring moments causes a uniform canting of the spins, leading to a weak ferromagnetism per CuO$_{2}$ plane, along with AFM ordering [Fig.~\ref{fig:CrystalStructure}(a)].  When La$_{2}$CuO$_{4}$ is doped with charge carriers, \textit{e.g.} via Sr or Li substitution, the carriers frustrate the magnetism, leading in both cases to a strong reduction of the N\'eel temperature $T_N$ \cite{rykov95,sarrao96}. Different types of magnetic textures may be stabilized depending on the geometry and character of the dopants.  For example, the diagonal incommensurate magnetic signatures reported by neutron-scattering experiments at very low (insulating) doping ($x$) \cite{matsuda} when Sr$^{2+}$ replaces La$^{3+}$ on top of a $4$-Cu plaquette can be explained well by the formation of local spin spirals \cite{shraiman90,sushkov}.  At higher $x$, the incommensurate magnetic signatures, now rotated by 45$^{\circ}$, have been consistently attributed to the emergence of spin and charge stripes \cite{kivelson}.
The substitution of Li$^+$ for Cu$^{2+}$ in-plane, on the other hand, may stabilize the formation of skyrmions \cite{haas96} that are obtained, roughly speaking, by reversing the average spins in a finite region of space around the dopant [Fig.~\ref{fig:CrystalStructure}(b)].  At low $x$, skyrmions are expected to affect only the tails of NMR or neutron line shapes \cite{haas96}, making any unambiguous interpretation difficult,
but even this has not been reported so far.  Here instead we take advantage of the weak FM moments to tune the orientation of the AFM background using a uniform $B$.  This gives rise to magnetotransport that is sensitive to the spin configuration around the dopants, and consequently to its nontrivial textures.  The resulting MR data on AFM La$_{2}$Cu$_{0.97}$Li$_{0.03}$O$_{4}$ are consistent with the presence of local skyrmions.

A high-quality La$_{2}$Cu$_{1-x}$Li$_{x}$O$_{4}$ (Li-LCO) single crystal with a nominal $x = 0.03$ was grown by the traveling-solvent floating-zone technique \cite{sasagawa02}. Magnetization $M$ was measured in a standard
superconducting quantum interference device magnetometer in $0.1\leq B$(T)$\leq 7$ applied either parallel or perpendicular to the CuO$_{2}$ ($ab$) planes. The in-plane resistance $R$ was measured on a bar-shaped sample with dimensions $2.2 \times 0.57 \times 0.41$ mm$^{3}$ using a standard four-terminal ac method ($\sim7$~Hz) in the Ohmic regime.  The contacts were made by evaporating Au and annealing at 700~$^{\circ}$C in air. MR was measured by sweeping either $B\parallel ab$ or $B\perp ab$ at constant $T$ in the 5-190~K range.  The sweep rates were low enough to avoid the heating of the sample due to eddy currents.

\begin{figure}

\includegraphics[width=4.0cm]{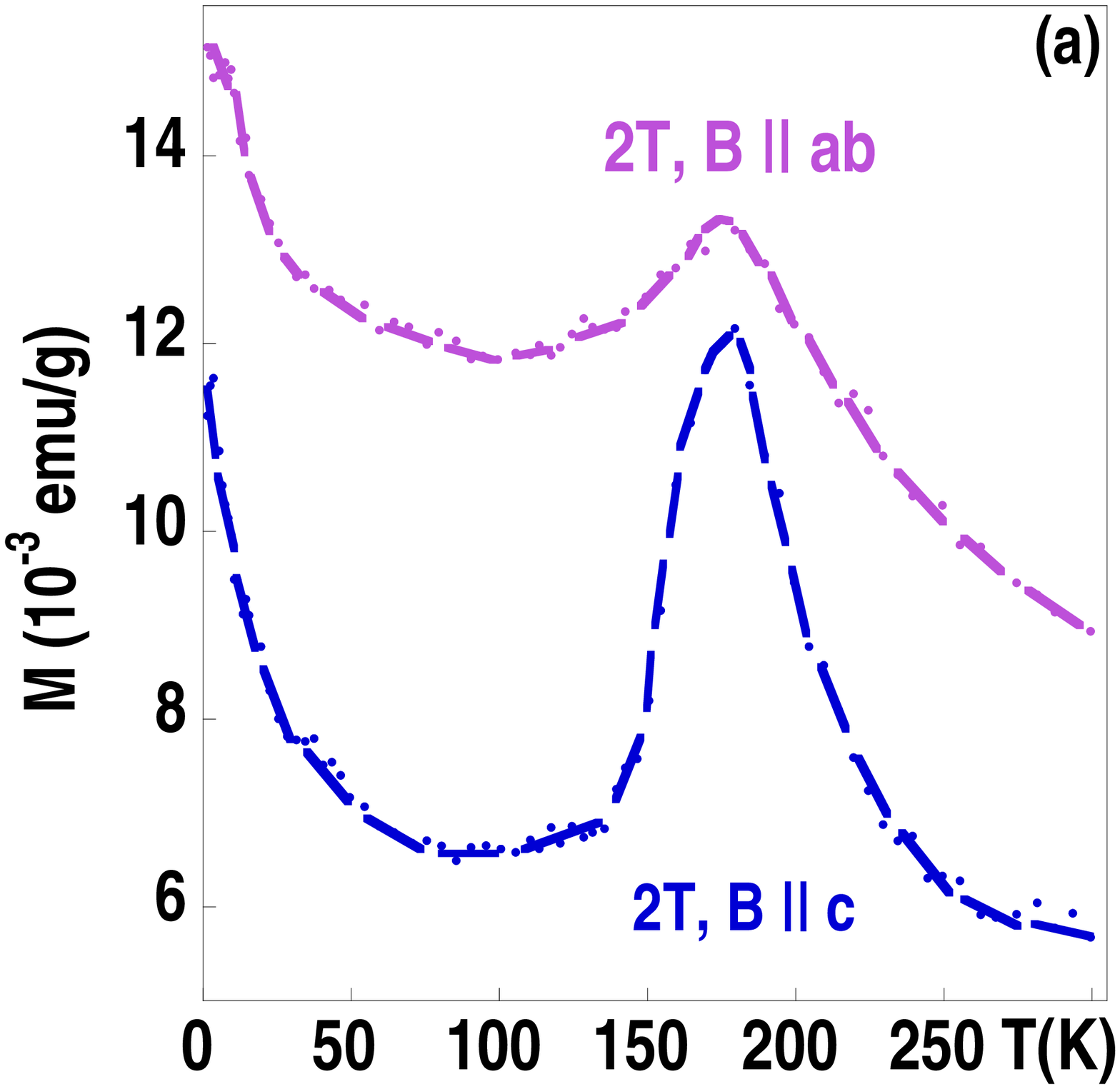}
\includegraphics[width=4.0cm]{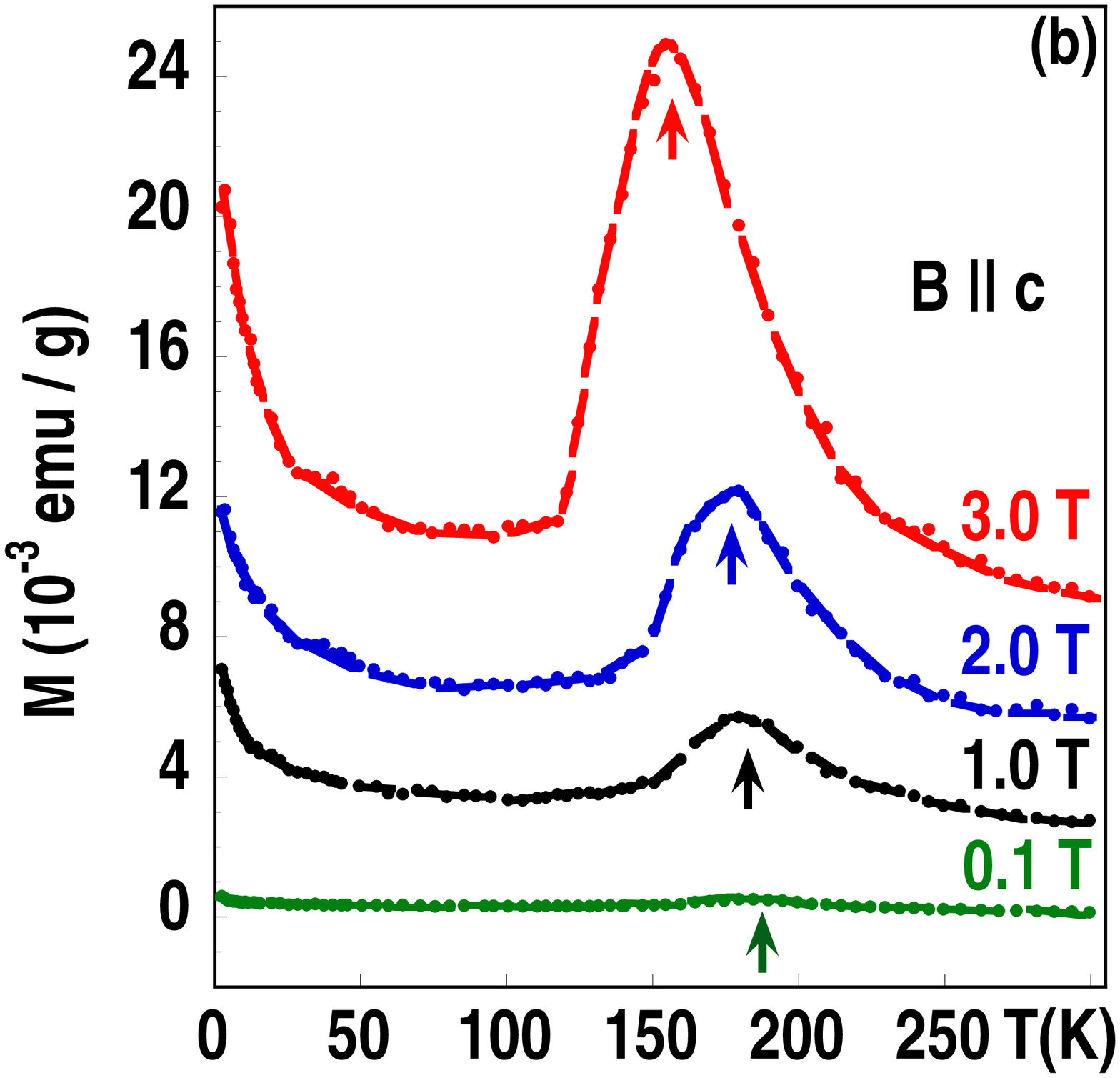}\\
\includegraphics[width=4.0cm]{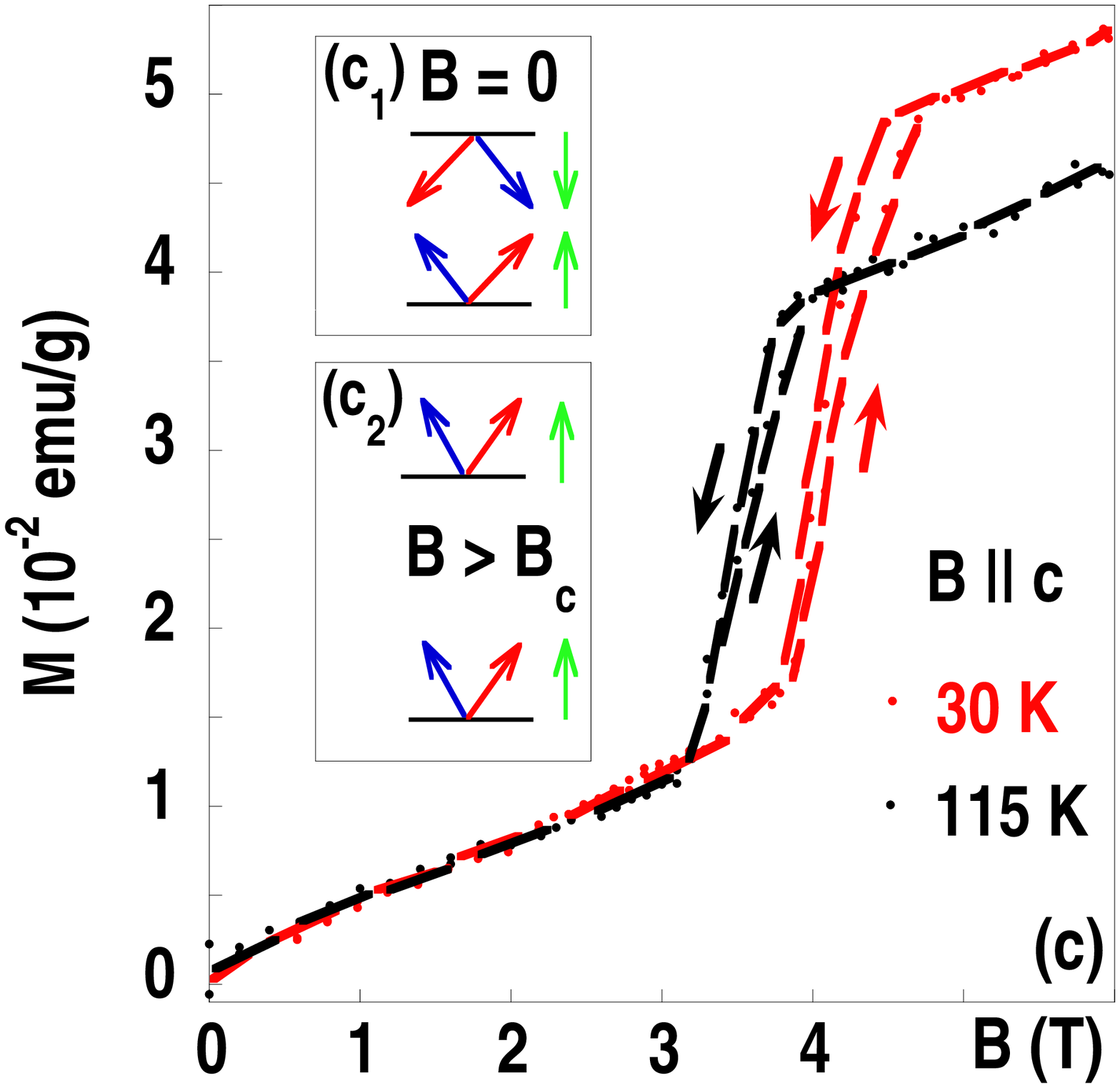}
\includegraphics[width=4.0cm]{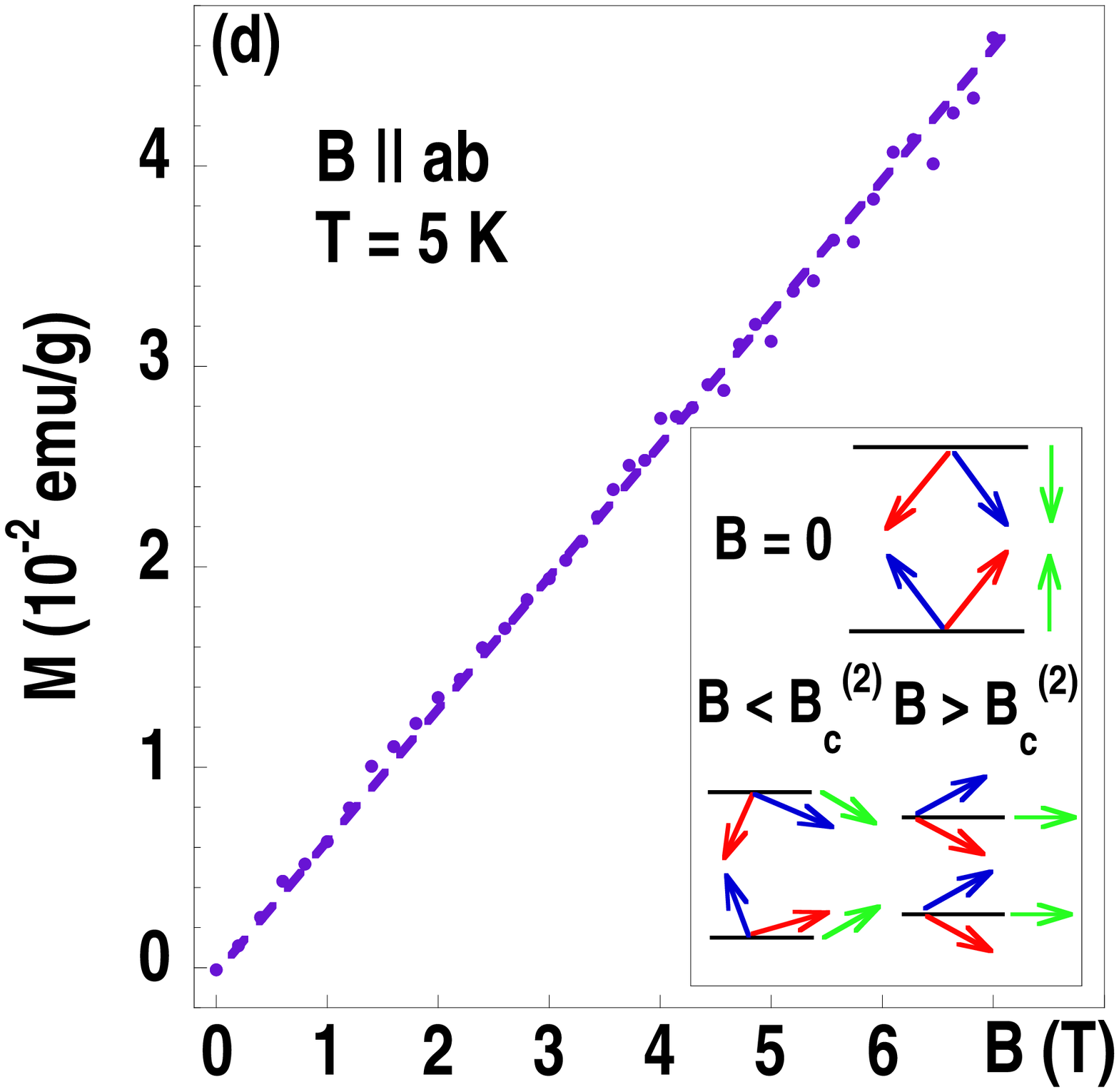}

\caption {(color online) (a) $M$ \textit{vs.} $T$ for $B\parallel c$ and $B\parallel ab$. The $B\parallel ab$ curve is shifted up by 3.5$\times 10^{-3}$ emu/g. (b) $M$ \textit{vs}. $T$ for several $B\parallel c$.  The curves at $1$~T and $0.1$~T are shifted down by 1.5$\times 10^{-3}$ emu/g and 1$\times 10^{-3}$ emu/g, respectively. Arrows denote the N\'{e}el temperature; $T_{N}(B=0)\approx 180$~K.  (c) $M$ \textit{vs.} $B\parallel$ c at different $T$. The arrows denote the direction of $B$-sweeps. Insets: Spin configuration for (c$_{1}$) $B = 0$ and (c$_{2}$) $B > B_{c}$, where $B_{c}$ is the critical field for the spin-flop transition (red and blue arrows: full Cu spins; green arrows: weak FM moments).  (d) $M$ \textit{vs.} $B\parallel ab$ at $T=5$ K.  Inset: Continuous rotation of spins in the $bc$ plane for $B\parallel b$, where $B_{c}^{(2)}$ is the saturation field.  Dashed lines guide the eye.}\label{fig:NeelTandMvsB}
\end{figure}

In La-based AFM systems, the easy axis for the spins is the longest ($b$) of the two in-plane orthorhombic directions \cite{benfatto06}.  The DM vector $\textbf{D}$ is oriented along $a$, the AFM order parameter $\textbf{n}_0$ along $b$, so that the weak FM moments $\textbf{L}= \textbf{D}\times\textbf{n}_0$ are parallel to $c$ [Fig.~\ref{fig:CrystalStructure}(a)].  One signature of the presence of the weak FM moments is a peak in the magnetic
susceptibility at $T_{N}$, such that it is more pronounced for $B\parallel c$ than for $B\parallel b$\, \cite{thio88,lavrovSS01}.  This is indeed observed also in Li-LCO [Fig.~\ref{fig:NeelTandMvsB}(a)].  Since the sample
is twinned, the $B$
%
%%%%%%%%%%%%%%%%%%%%%%%%%%%%%%%%%%%%%%%%%%%%%%%%%%%%%%%%%%%%%%%%%%%%%%%%FIG2
%
%%%%%%%%%%%%%%%%%%%%%%%%%%%%%%%%%%%%%%%%%%%%%%%%%%%%%%%%%%%%%%%%%%%%%%%%%%%%%%%%%%%%%%%%%%%%%%%%%%%%%%FIG2
%
direction within the plane is not specified, but only twins having $B\parallel b$ contribute to anomalies in $M$ (and resistivity $\rho$).  $T_{N}$ is suppressed by both $B\parallel c$ [Fig.~\ref{fig:NeelTandMvsB}(b)] and $B\parallel ab$ (not shown).  Moreover, the low-$T$ upturn of $M$ is similar to that in undoped and Sr-doped La$_2$CuO$_4$ (LSCO) \cite{lavrovSS01}, and thus it is irrelevant to the skyrmion formation. A second signature of the weak FM moments is found in $M(B)$ at fixed $T$.  Because of the weak interplane AFM coupling, the weak FM moments have a staggered order along the $c$ axis [Fig.~\ref{fig:NeelTandMvsB}(c) inset (c$_{1}$)].  A sufficiently large field $B_{c}\parallel c$ can overcome the interplane AFM coupling and induce a discontinuous spin-flop reorientation in both the weak FM and the large AFM components [Fig.~\ref{fig:NeelTandMvsB}(c) inset (c$_{2}$)], causing the so-called weak FM (WFM), first-order transition \cite{thio88,andoMR03,benfatto06}. This results in a jump $\Delta M(T)$ at $B_{c}(T)$ \cite{thio90,andoMR03} [Fig.~\ref{fig:NeelTandMvsB}(c)].  Both $\Delta M(T)$ and $B_{c}(T)$ decrease with $T$
\cite{thio88,thio90,andoMR03}, following the decrease of the staggered magnetization n$_{0}$(T) due to thermal fluctuations \cite{benfatto06}.  For $B\parallel b$, the weak FM moments induce a continuous rotation of $\textbf{n}_{0}$ in the $bc$ plane \cite{benfatto06,andoMR04} (Fig.~\ref{fig:NeelTandMvsB}(d) inset), so that $M(B)$ increases smoothly [Fig.~\ref{fig:NeelTandMvsB}(d)].  Thus the magnetic properties of Li-LCO resemble very much those of Sr- and O-doped La$_2$CuO$_4$ \cite{thio88,thio90,lavrovSS01} where skyrmions cannot form.

\begin{figure}

\includegraphics[width=4.1cm]{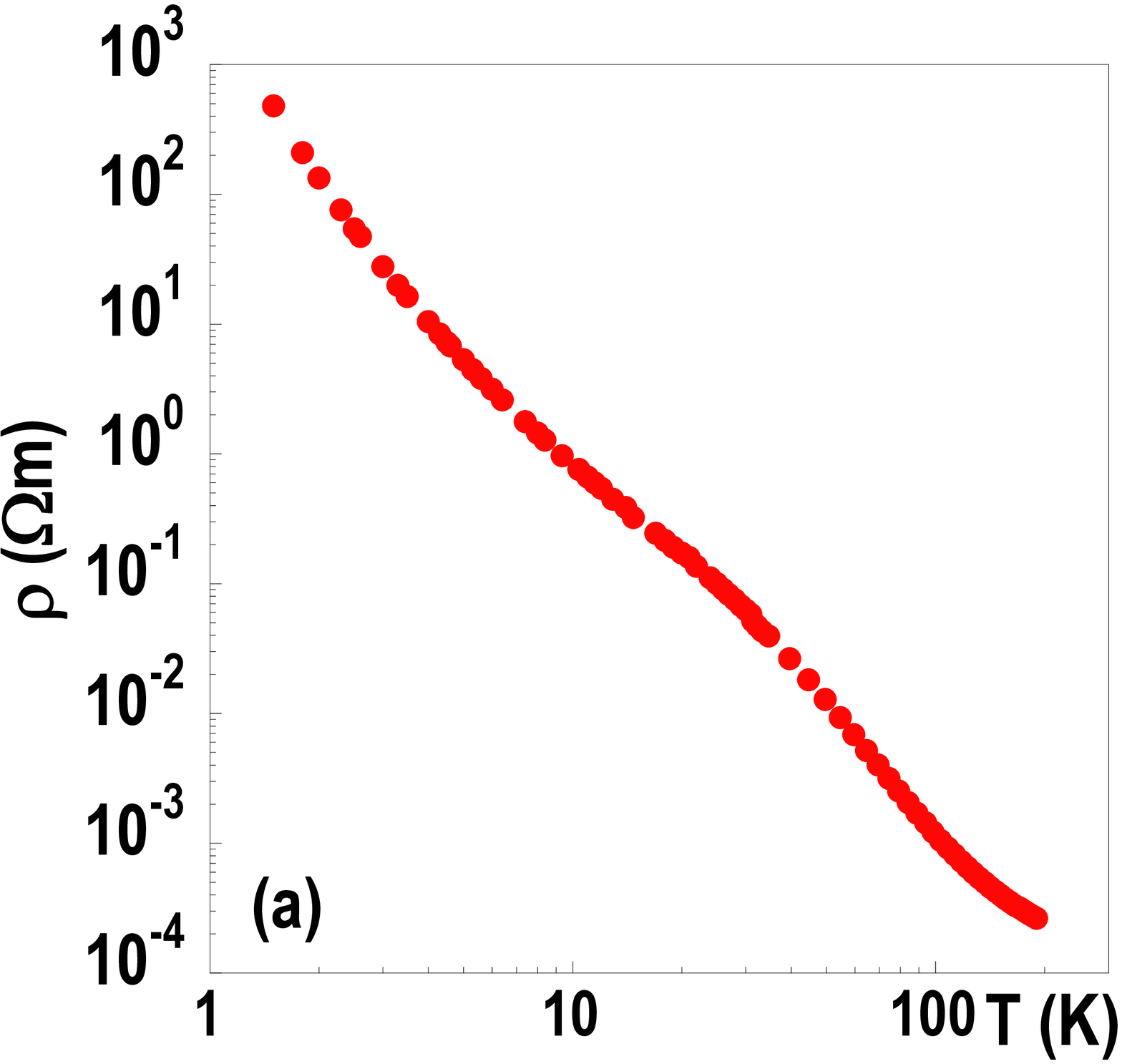}
\includegraphics[width=4.1cm]{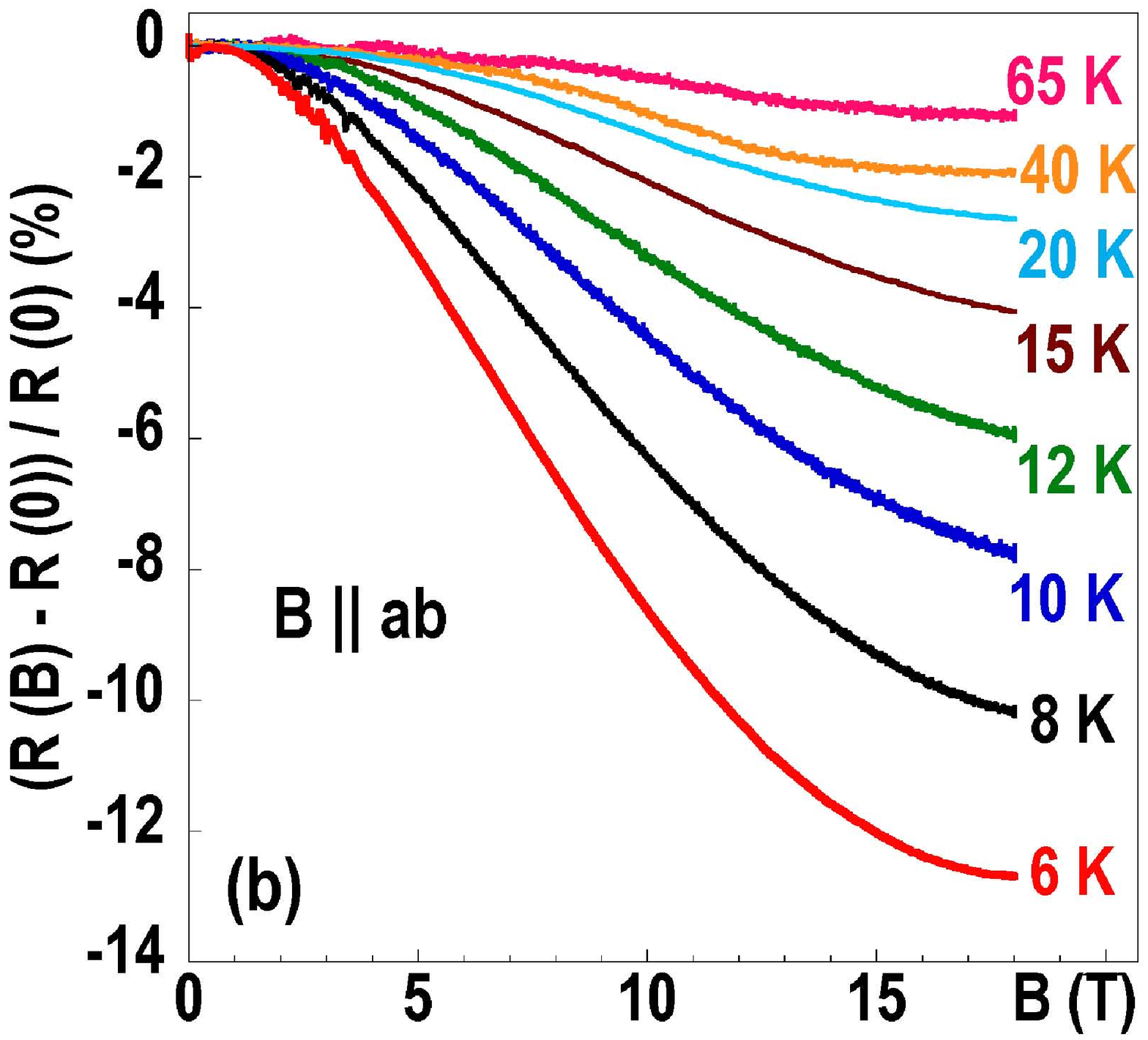}\\
\includegraphics[width=4.0cm]{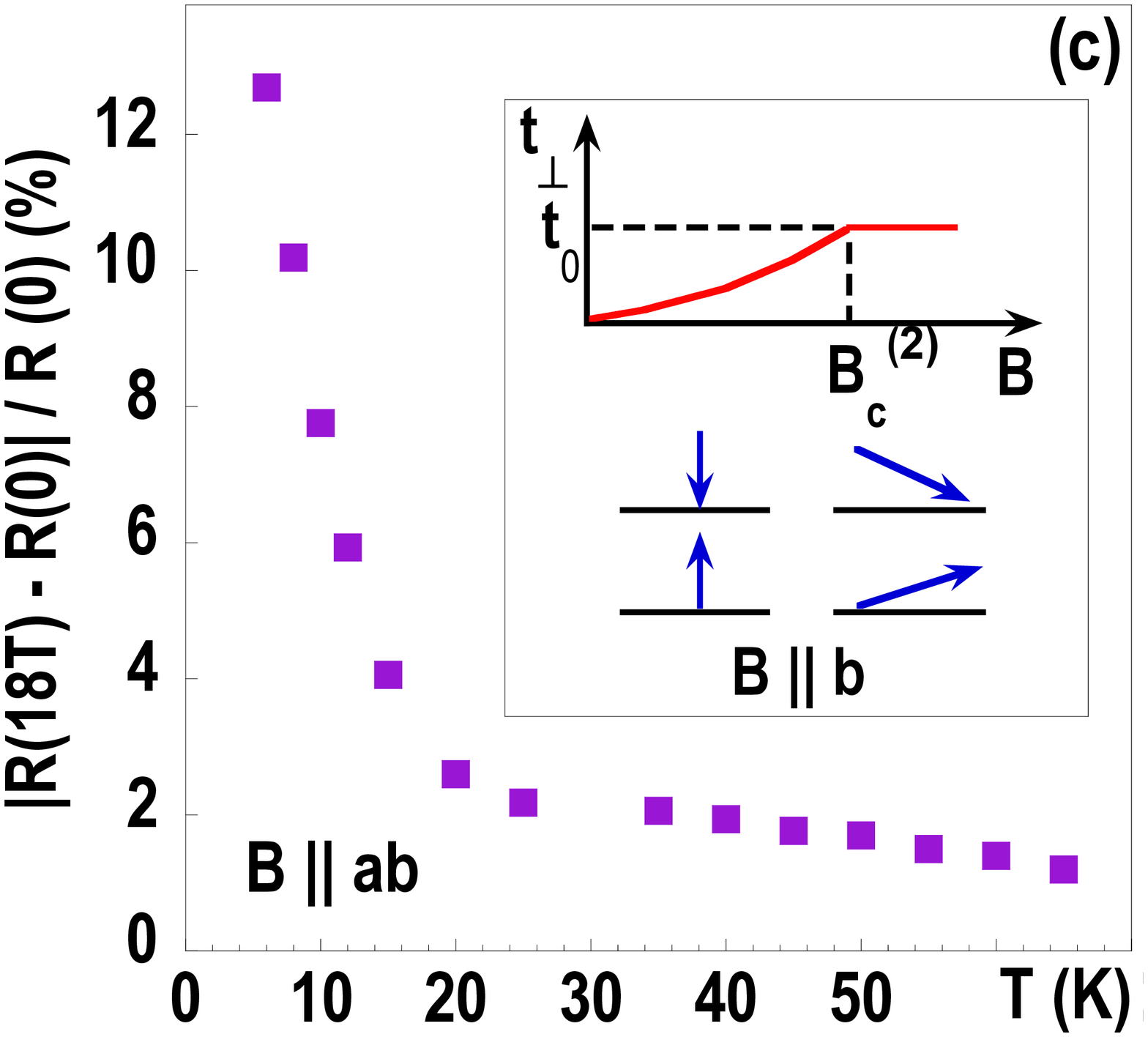}
\includegraphics[width=4.0cm]{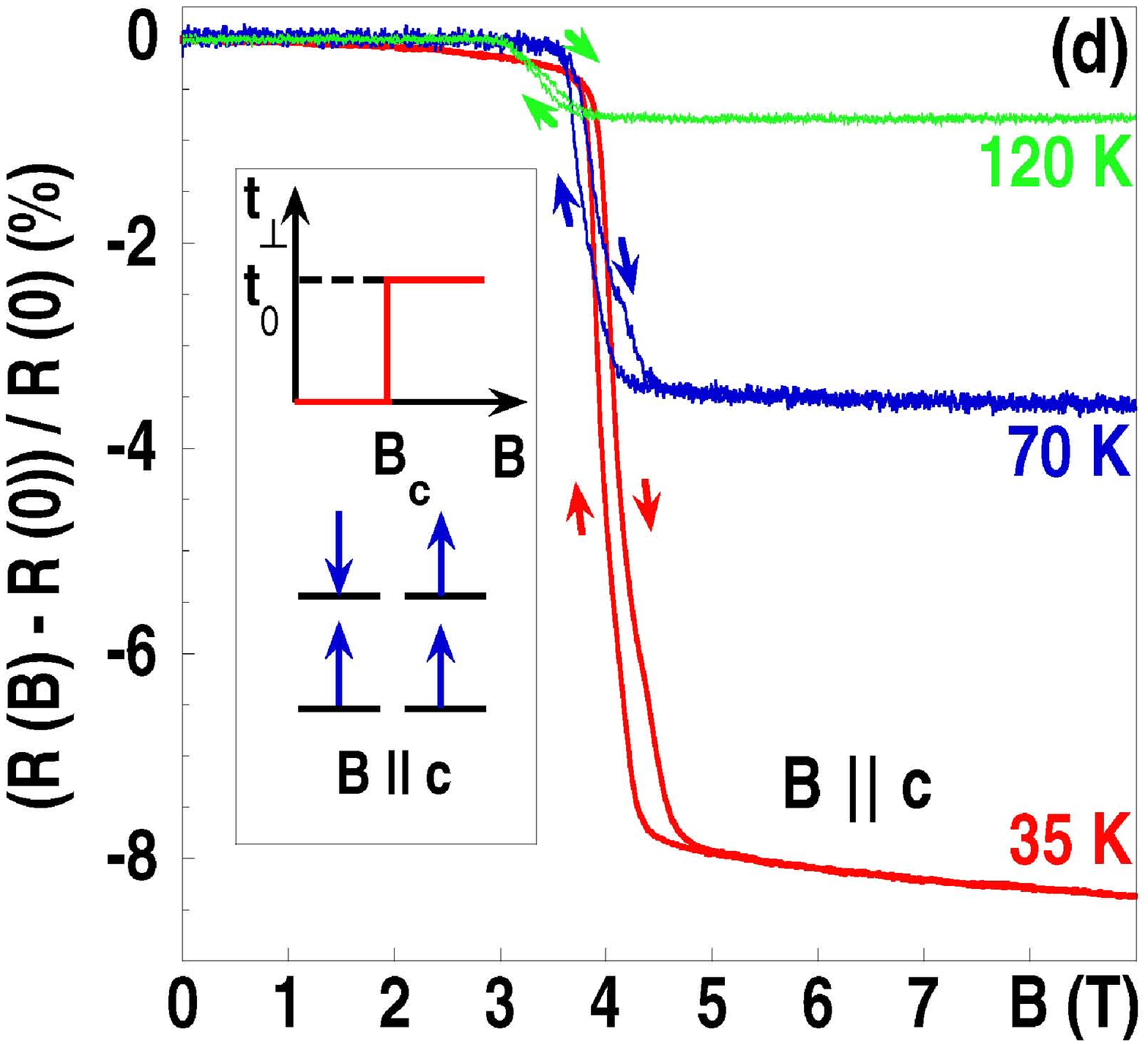}\\
\includegraphics[width=4.1cm]{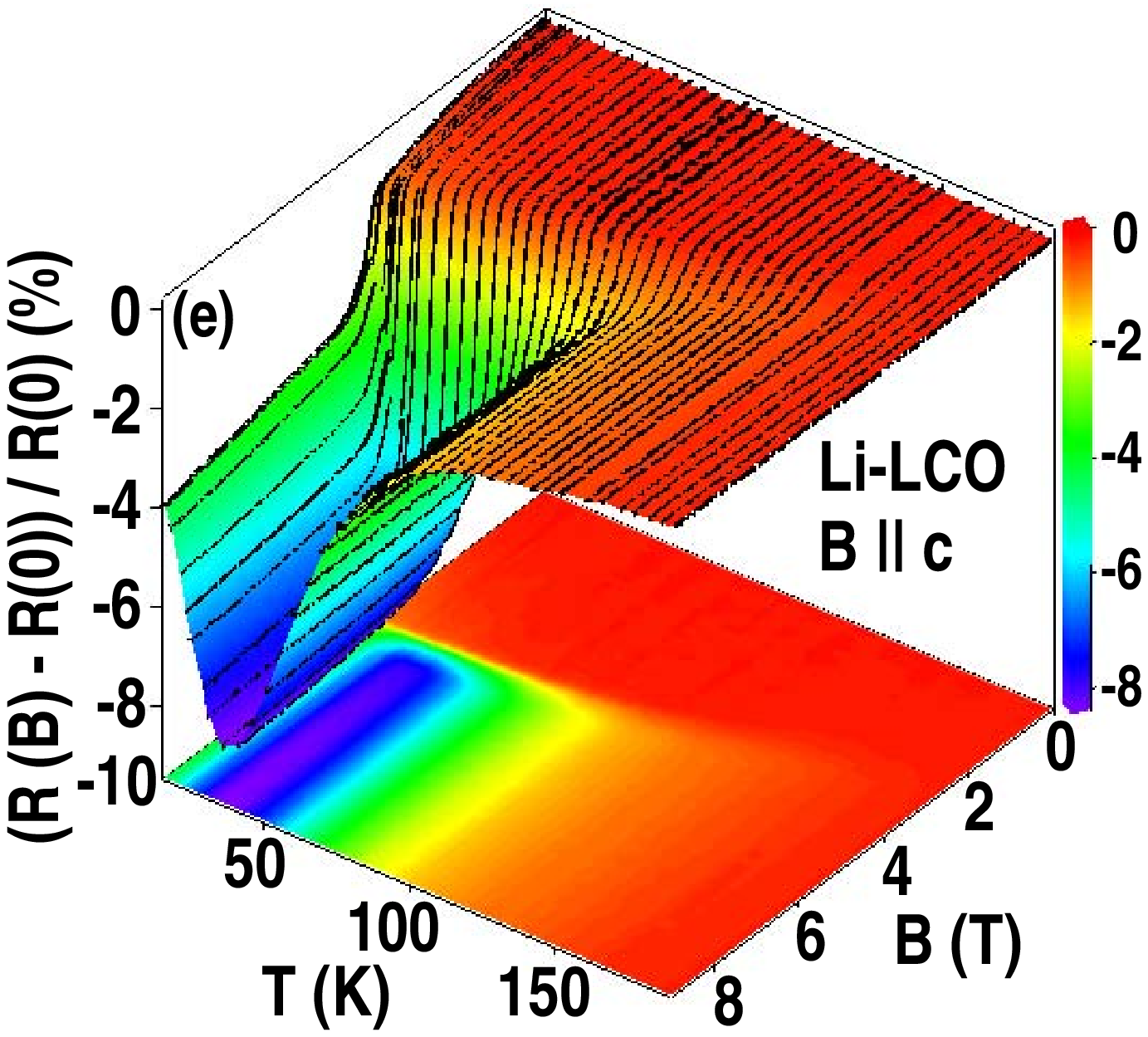}
\includegraphics[width=4.4cm]{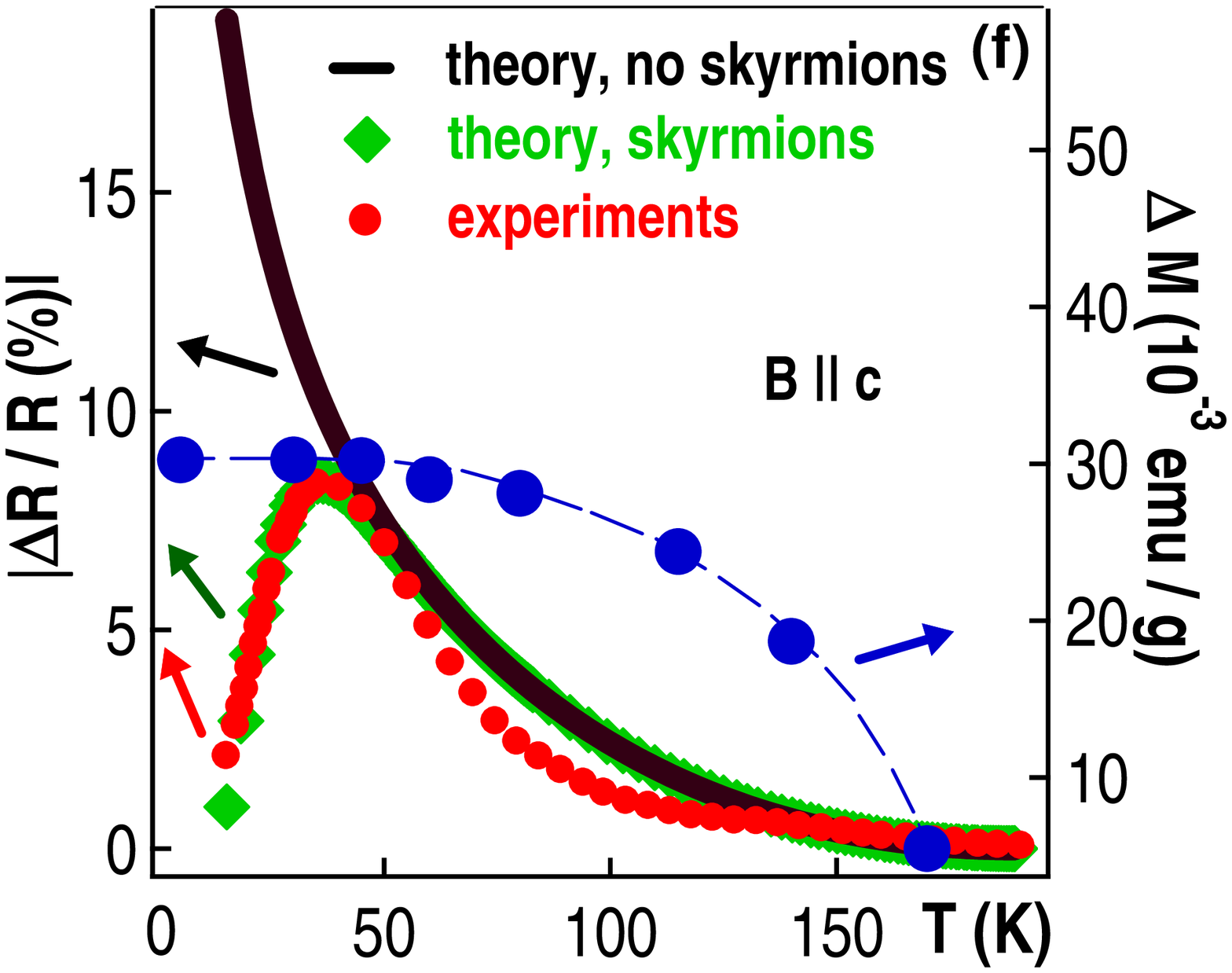}

\caption{(color online) (a) $\rho$ \textit{vs.} $T$ at $B=0$.  (b) MR for $B\parallel ab$ at several $T$.  (c) MR \textit{vs.} $T$ at 18~T for $B\parallel ab$.  Inset: The sketch of the weak FM moments (blue arrows) and the corresponding interlayer hopping $t_\perp$ for $B\parallel ab$ ($t_{0}$, maximum value of $t_\perp$; $B_{c}^{(2)}$, saturation field).  (d) MR for $B\parallel c$ at three selected $T$. Arrows denote the directions of the $B$-sweeps.  Inset: Weak FM moments (blue arrows) before and after the weak FM transition at $B_{c}$ and the corresponding field-induced $t_{\perp}$.  (e) MR for $B\parallel c$ at $15\leq T$(K)$\leq 190$.  (f) The nonmonotonic $T$ dependence of the MR step size shown in (e), and the monotonic decrease of the magnetization step with $T$. The dashed line guides the eye.  The calculations of the MR step size in the presence and in the absence of skyrmions \cite{Suppl} are also shown.}\label{fig:Magnetotransport}
\end{figure}

At very low $x$, the motion of the holes may be described as the hopping of (spinless) charges in the two AFM sublattices \cite{kane89,shraiman90}.  Because of orthorhombicity, the majority of holes have momenta close to $(\pi/2,-\pi/2)~$\cite{kotov07}, \textit{i.e.} the most relevant pocket is the one along the $b$ direction. When the Coulomb potential provided by the dopant is taken into account, the holes get localized, with a typical 2D pancake-like exponential envelope $\psi (r)\sim \exp({-r/\xi_{0}})$ ($r$, planar distance from the impurity site; $\xi_{0}$, localization length) \cite{kotov07}.  The low-$T$ electronic transport is then expected to occur via variable-range hopping (VRH) between localized states \cite{efros} at a characteristic hopping distance $r_{h}(T)$.  Indeed, the insulating behavior of $R$ [Fig.~\ref{fig:Magnetotransport}(a)] follows a 2D VRH form $R\propto \exp({T_{0}/T})^{1/3}$ ($T_0=3753$~K) for $T<23$~K, and it crosses over to $R\propto \exp({E_{A}/T})$ ($E_A=313$~K) at higher $T$ \cite{Suppl}, similar to the behavior of other La-based AFM samples \cite{kastner98}.

%%%%%%%%%%%%%%%%%%%%%%%%%%%%%%%%%%%%%%%%%%%%%%%%%%%%%%%%%%%%%%%%%%%%%%%%FIG3

%%%%%%%%%%%%%%%%%%%%%%%%%%%%%%%%%%%%%%%%%%%%%%%%%%%%%%%%%%%%%%%%%%%%%%%%%%%%%%%%%FIG3

The in-plane MR $[R(B)/R(0)-1]$ at first glance also resembles that of AFM LSCO \cite{andoMR03,andoMR04}.  For example, for $B\parallel ab$, the MR is negative, decreases continuously with a tendency to saturate above some threshold field [Fig.~\ref{fig:Magnetotransport}(b)], and its overall magnitude increases monotonically as $T$ is reduced [Fig.~\ref{fig:Magnetotransport}(c)].  For $B\parallel c$, MR is also negative, with a steplike decrease [Fig.~\ref{fig:Magnetotransport}(d)] at the same critical field B$_{c}$ where the WFM transition occurs and the uniform $M$ shows a jump [Fig.~\ref{fig:NeelTandMvsB}(c)].  Moreover,
both $M$ [Fig.~\ref{fig:NeelTandMvsB}(c)] and MR [Fig.~\ref{fig:Magnetotransport}(d)] exhibit a hysteresis associated with a first-order phase transition.  However, as shown below, it is the nonmonotonic $T$ dependence of the MR magnitude [Fig.~\ref{fig:Magnetotransport}(e)] and $|\Delta R|/R$ [Fig.~\ref{fig:Magnetotransport}(f)], the MR step size at $B_c$, that signals the formation of skyrmions in the ground state of AFM Li-LCO.  At high $T$, $|\Delta R|/R$ grows with decreasing $T$ as expected, but unlike AFM LSCO \cite{andoMR03}, it exhibits a dramatic reversal of the behavior below $T\approx T_{S} = 35$~K, and starts to decrease as $T$ is reduced further.  We attribute this decrease to the progressive emergence of skyrmions of increasing core size that suppress the interlayer tunneling process responsible for the negative MR in doped AFM La$_2$CuO$_4$ \cite{kotov07}.

In general, holes can only tunnel between planes along the direction of FM order of the spins: $b$ at $B=0$ [Fig.~\ref{fig:CrystalStructure}(a)] and $a$ above the WFM transition. However, since the hole momentum is close to $(\pi/2,-\pi/2)$ due to orthorhombicity, this hopping process is suppressed at $B=0$.  It becomes possible only above the WFM transition \cite{Suppl}, modifying the in-plane wave function as $\psi_{B}(r)=f(r,B)\psi_{0}(r)$ ($\psi_{0,B}$ are the wave functions at zero and finite $B$, respectively).  In the hopping-conductivity scheme described above, $R\sim\mid\psi(r_{h})\mid^{-2}$, so that the MR is given by $ \frac{R(B)-R(0)}{R(0)}= -(1-\frac{\mid\psi_{0}(r_{h})\mid^{2}}{\mid\psi_{B}(r_{h})\mid^{2}})$.  At the typical length scale $r_h(T)$ for hopping, $f(r_h,B)\simeq 1+\alpha~(r_h\,t_{\perp}(B))^{2}> 1$~\cite{kotov07}, leading to a negative MR, as observed in Figs.~\ref{fig:Magnetotransport}(b) and \ref{fig:Magnetotransport}(d)-(e).  Here $t_{\perp}$ is the interlayer hopping and $\alpha=2/(\xi_0\epsilon_0)^2$ ($\epsilon_0$, the localization energy of the hole bound state).  When $B\parallel b$, $t_{\perp}(B)= t_{0}\sin\theta(B)$ increases continuously with $B$
(Fig.~\ref{fig:Magnetotransport}(c) inset); here $\theta(B)$ is the angle with the $b$ direction that spins form in their smooth rotation in the $bc$ plane (see also Ref. \cite{andoMR04}). However, for $B\parallel c$, $t_{\perp}(B)$ jumps discontinuously (Fig.~\ref{fig:Magnetotransport}(d) inset) from zero to the maximum value $t_{0}$ at the critical field $B_{c}$.  Hence, the MR is continuous \cite{andoMR04} for $B\parallel b$ [Fig.~\ref{fig:Magnetotransport}(b)], while it is discontinuous \cite{thio90,andoMR03} for $B\parallel c$ [Fig.~\ref{fig:Magnetotransport}(d)].

In this picture, the $T$ dependence of the step size $|\Delta R|/R$ is determined by $r_h(T)$ and $t_\perp(T)$.  As the transport crosses over from the activated to the VRH regime (where $r_h(T)\sim\xi_0(T_0/T)^{1/3}$) at low $T$, $r_h(T)$ increases with decreasing $T$, and one expects only a monotonic increase of the MR magnitude.  This is indeed observed for $B\parallel b$ [Fig.~\ref{fig:Magnetotransport}(c)] and in AFM LSCO \cite{andoMR03}.  Since the MR behavior is intimately related to the AFM order, any mechanism that leads to a global reduction of the AFM order parameter n$_{0}(T)$ may be expected to contribute also to a suppression of $|\Delta R|/R$.  However, the low-$T$ drop of $|\Delta R|/R$ cannot be ascribed to any such mechanism, since the magnetization step $\Delta M(T)$, which is proportional to n$_{0}(T)$ \cite{benfatto06}, increases continuously in the same $T$ range [Fig.~\ref{fig:Magnetotransport}(f)].  This rules out, for example, spin glassiness, which is known to reduce $\Delta M(T)$ in AFM LSCO \cite{yamada}, but which here sets in only below $7-8$~K \cite{sasagawa02}.  It also rules out a structural change of the kind observed in La$_{1.79}$Eu$_{0.2}$Sr$_{0.01}$CuO$_4$ \cite{huecker}, although none have been reported on Li-LCO in our $T$ range.  Charge glassiness, which gives rise to a positive MR, is also not relevant, as it sets in at much lower $T$ \cite{Ivana-pMR}.  Thus the only mechanism that can lead to
a decrease in $|\Delta R|/R$ is the local suppression of $t_\perp$, as it happens when a skyrmion forms.

Close to the Li impurity position $r = 0$, where the skyrmion is centered, ${\bf n}_0$ is nearly fully reversed with respect to the equilibrium direction and recovers only at distances $r\geq \lambda$
[$\lambda$, the characteristic length scale fixing the core size of the skyrmion, Fig.~\ref{fig:CrystalStructure}(c)].
At $B=0$, the ordering of spins in the two neighboring layers, the one with the impurity and the one above it, is nearly FM within a distance of the order of $\lambda$.  For $B>B_{c}$, a full reversal of both weak FM and staggered moments occurs: the ordering of the magnetic moments within a distance $\lambda$ from the impurity is AFM
[Fig.~\ref{fig:CrystalStructure}(c)], and $t_\perp$ is suppressed for distances up to $r\simeq \lambda$.
Thus, if the formation of skyrmions suppresses $t_\perp$ in $f(r_h,B)$, the MR step size is expected to decrease.
By modeling the gradual increase of $\lambda$ as $T$ is lowered \cite{Suppl}, we have reproduced [Fig.~\ref{fig:Magnetotransport}(f)] the observed nonmonotonic behavior in $|\Delta R|/R$ for $B\parallel c$, characterized by a pronounced downturn below $T_{S} = 35$~K. This behavior is contrasted to the smooth and monotonic increase in the case where no skyrmions are formed \cite{kotov07}.  Since $\lambda$
obtained in our analysis is of the order of one lattice spacing \cite{Suppl}, such nearly pointlike skyrmions will interact very weakly among themselves and they are not expected to
cause significant changes in bulk quantities, such as $M$.
They will be crucial, however, in suppressing local processes, such as the vertical tunneling around the impurity, associated with the negative MR effect.

Skyrmions carry a nonzero topological charge, $Q\neq 0$, and a skyrmion configuration is orthogonal to the $Q=0$ N\'eel state. In order to stabilize a skyrmion at $T=0$, the hole state must modify the topology of the CuO$_2$ layers.  Since the spins on the four nearest neighbors to the Li sites belong to the same AFM sublattice
[Fig.~\ref{fig:CrystalStructure}(a)], the localized hole wave function can be labeled by the eigenvalues of the orbital angular momentum: $\ell=\pm 1$ (excited states) or $\ell=\pm \rm i$ (lowest energy states) \cite{gooding91}. The $\ell=\pm \rm i$ eigenstates have circulation associated with them, describing a localized hole orbiting around the Li center \cite{gooding91}.  The associated,  nonzero spin current induces the skyrmion deformation of the AFM background \cite{gooding91,haas96}.  This circulation might, or might not, be affected by an applied $B$, depending on whether the original symmetry of the problem is preserved.  For $B\parallel c$, the symmetry is preserved, $\ell=\pm \rm i$ are still good quantum numbers, and skyrmions survive the applied $B$.  For $B\parallel ab$, on the other hand, the symmetry is broken, the $\ell=\pm \rm i$ states
become mixed, causing the quenching of the hole angular momentum, and eventually skyrmion formation is suppressed. As a result, $|\Delta R|/R$ increases monotonically as $T$ is lowered [Fig.~\ref{fig:Magnetotransport}(c)], as expected without skyrmion formation [see theory curve in Fig.~\ref{fig:Magnetotransport}(f)]. Such a monotonic increase is indeed observed in both $B\parallel ab$ and
$B\parallel c$ for Sr doping \cite{andoMR03}, where skyrmions cannot occur since the hole orbital angular momentum is always quenched, from the start, by the mixing between the two sublattices around the position of the Sr dopant. For vanishingly small $B\parallel ab$, however, skyrmions might still be favorable and induce a nonmonotonic behavior for $|\Delta R|/R$, but probably for $T$ much lower than those accessed in our experiment.
Thus, what matters for the skyrmion formation is the local structure of the hole wave function at the Sr or Li sites, not the way transport occurs.  Indeed, in the AFM phase of interest, both LSCO \cite{andoMR04} and Li-LCO show similar VRH or activated behavior, with differences emerging only at higher doping \cite{sarrao96,rykov95}.

In summary, the low-$T$ magnetic and transport properties of the AFM La$_2$Cu$_{1-x}$Li$_x$O$_4$ provide the first experimental support for the predictions of skyrmions in AFM insulators.  Our work may offer new insights into the mechanisms that can stabilize or suppress topological excitations in complex magnetic systems.

We thank X. Shi for technical help,  V. Dobrosavljevi\'c, J. Lorenzana, A. N. Bogdanov for discussions, NSF DMR-0403491 and DMR-0905843, NHMFL via NSF DMR-0654118, MEXT-CT-2006-039047, EURYI, Italian MIUR Project PRIN-2007FW3MJX, and the National Research Foundation, Singapore for financial support.

\clearpage

\onecolumngrid
\begin{center}
\textbf{\large Supplementary material for ``Evidence for Skyrmions in a Doped Antiferromagnet''}\\
\vspace*{12pt}

I. Rai\v{c}evi\'{c},$^{1,2}$ Dragana Popovi\'c,$^{1,2}$ C. Panagopoulos,$^{3,4}$\\
L.~Benfatto,$^5$ M. B. Silva Neto,$^6$ E. S. Choi,$^1$ T. Sasagawa$^7$\\
\vspace*{6pt}

\textit{\small $^1$ National High Magnetic Field Laboratory, Florida State University, Tallahassee, FL 32310, USA\\
$^2$ Department of Physics, Florida State University, Tallahassee, FL 32306, USA\\
$^3$ Department of Physics, University of Crete and FORTH,71003 Heraklion, Greece\\
$^4$ Division of Physics and Applied Physics, Nanyang Technological University, Singapore\\
$^5$ CNR-ISC and Department of Physics, ``Sapienza'' University of Rome, 00185 Rome, Italy\\
$^6$ Instituto de F\' \i sica, Universidade Federal do Rio de Janeiro, CEP 21945-972, Rio de Janeiro - RJ, Brasil\\
$^7$ Materials and Structures Laboratory, Tokyo Institute of Technology, Kanagawa 226-8503, Japan}
\end{center}
\vspace*{9pt}

\twocolumngrid

The insulating behavior of the in-plane resistance $R$ of the antiferromagnetic La$_{2}$Cu$_{0.97}$Li$_{0.03}$O$_{4}$ sample [Fig.~3(a)] follows a 2D VRH form $R\propto \exp({T_{0}/T})^{1/3}$ for $T<23$~K (Fig.~S1 left).  At higher $T$, it crosses over to an activated law $R\propto \exp({E_{A}/T})$ (Fig.~S1 right).

%%%%%%%%%%%%%%%%%%%%%%%%%%%%%%%%%%%FIG-S1
\begin{figure}[h]

\includegraphics[width=4.0cm]{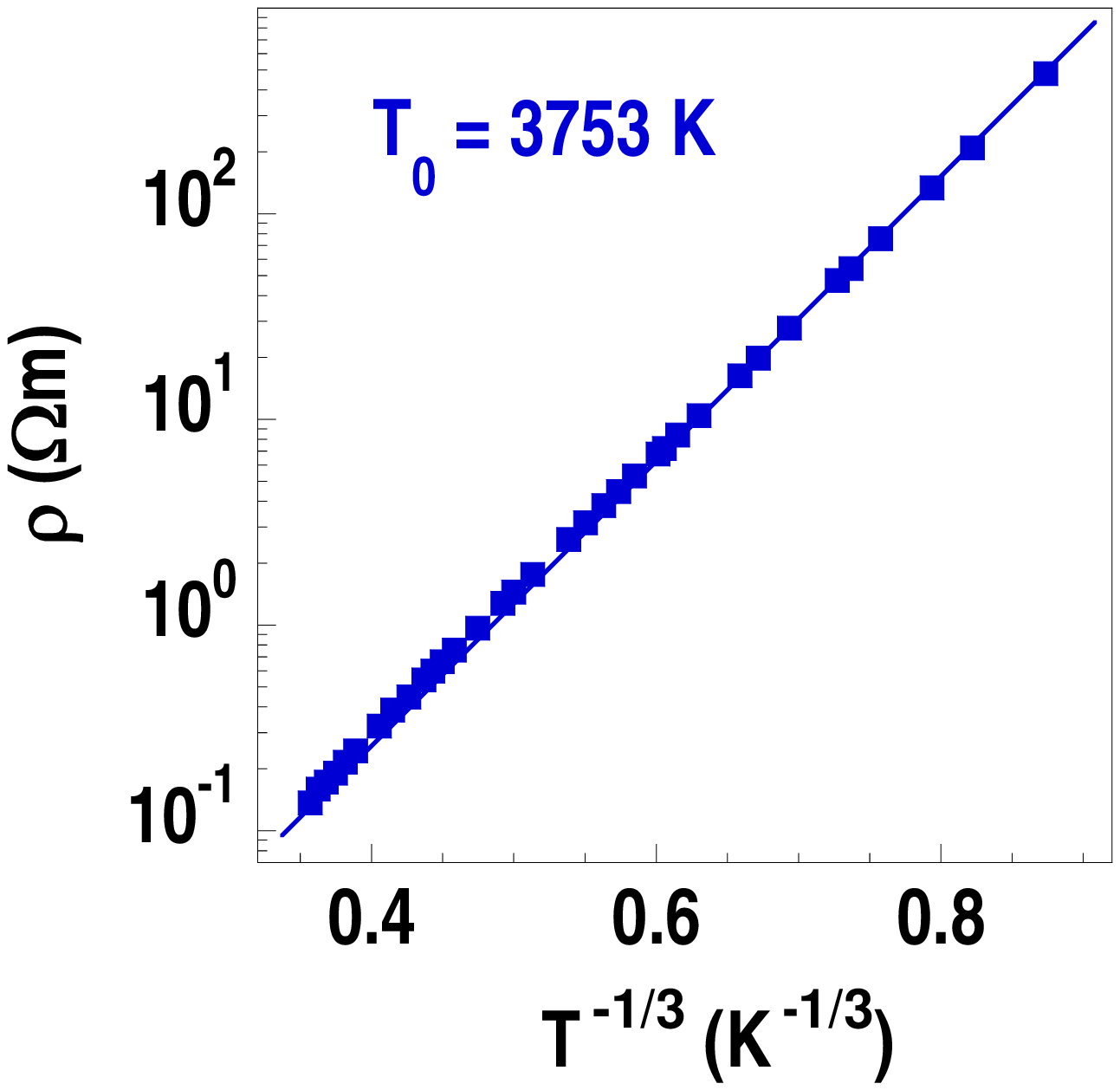}
\includegraphics[width=4.0cm]{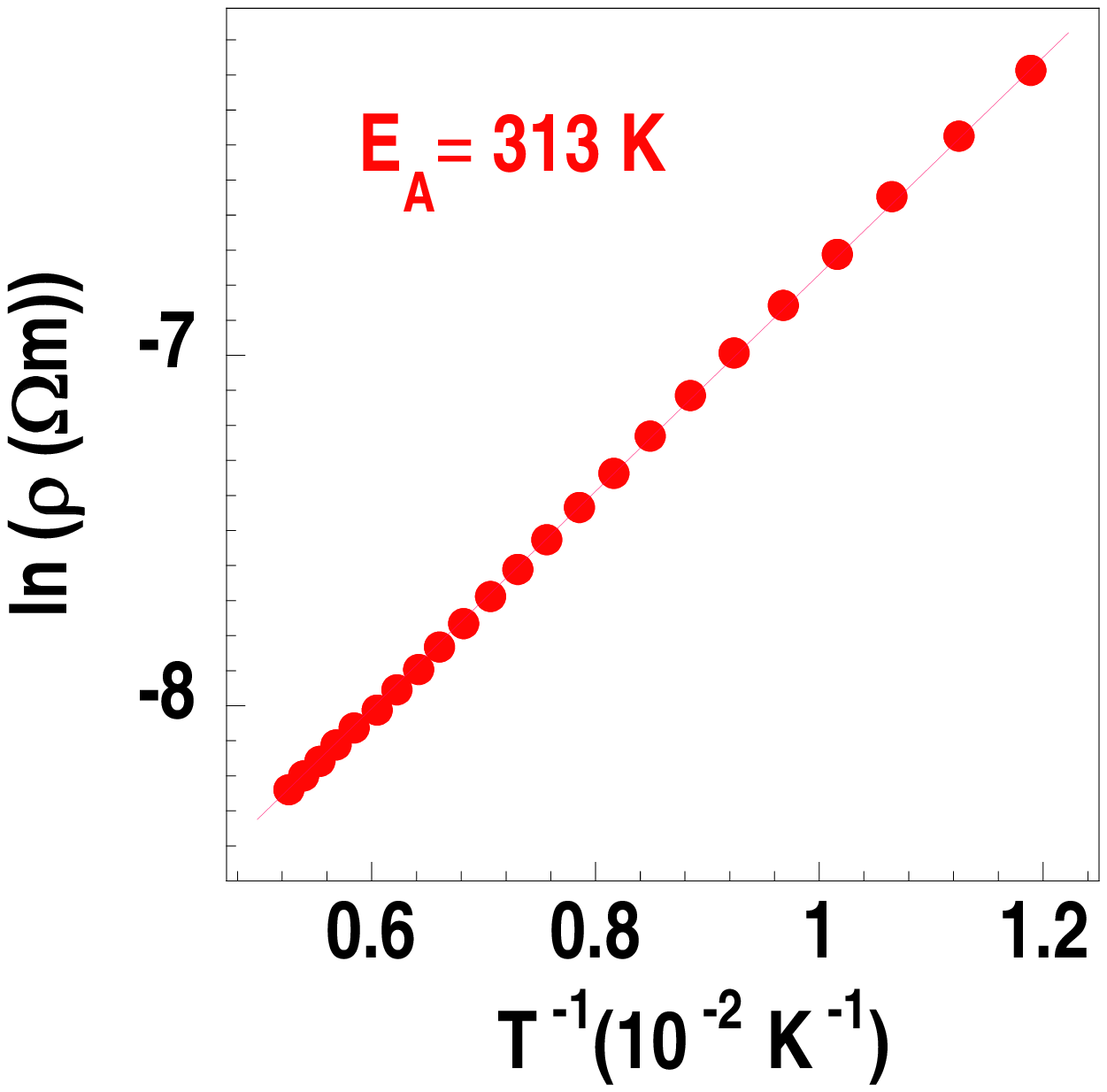}

\end{figure}
\noindent {\small FIG. S1 (color online) Left: 2D VRH behavior of $\rho (T)$ for $1<T$(K)$\leq 23$.  Right: Activated behavior of $\rho (T)$ for $80\leq T$(K)$\leq 190$.}
\vspace{12pt}
%
%%%%%%%%%%%%%%%%%%%%%%%%%%%%%%%%%%%%%%%%%%%%%%%%%%%%%%%%%%%%%%%%%%%%%%%%%%%%%%%%%%%%%%%%%%%%%%%%%%%%%%FIG-S1

Figure S2 shows the weak FM transition in the magnetization $M$ in $B \parallel c$ [Fig. 2(c)] at several temperatures and in magnetic fields all the way up to 7~T.

%
%%%%%%%%%%%%%%%%%%%%%FIG-S2
\begin{figure}[h]
\includegraphics[width=6.0cm]{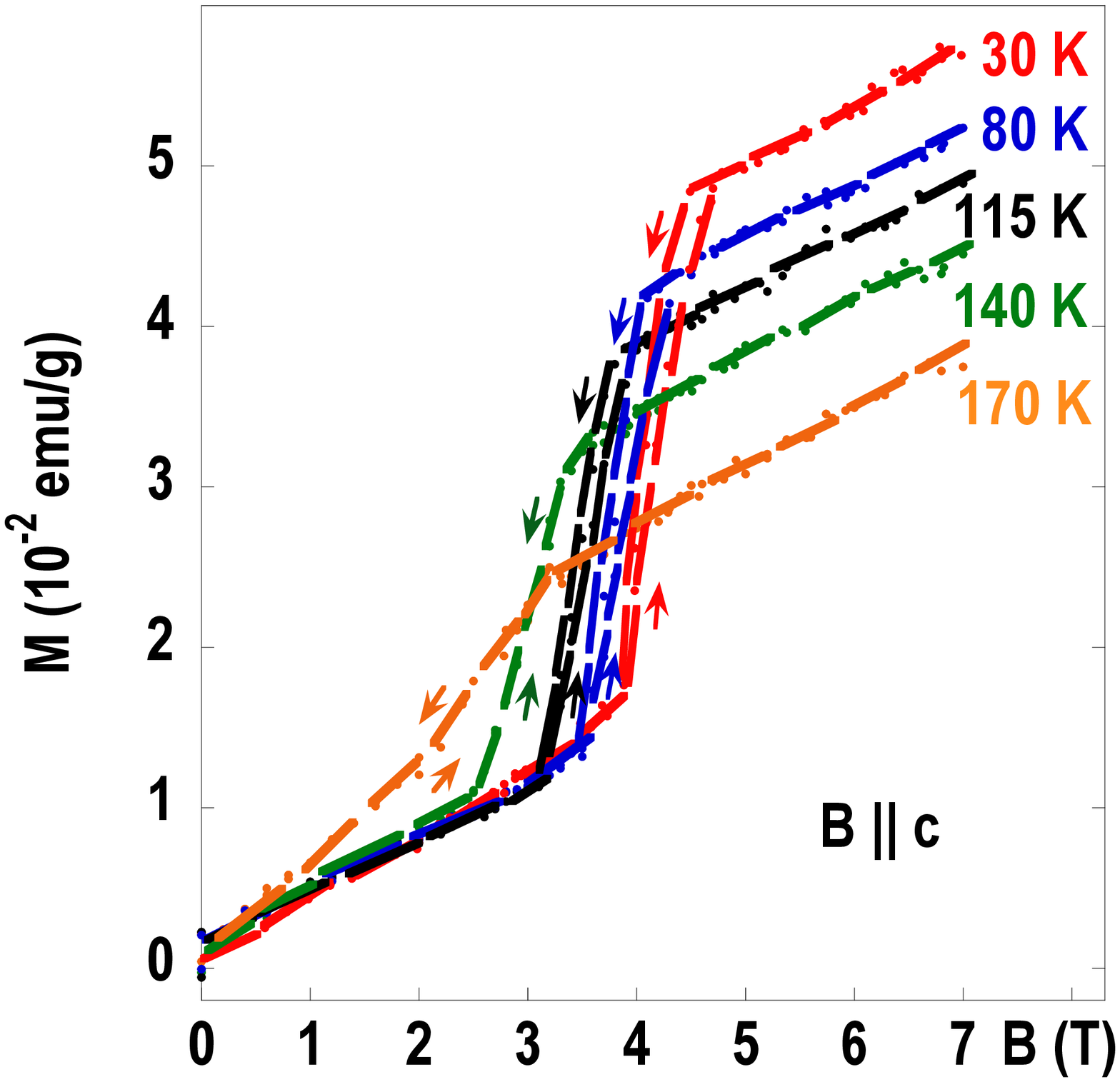}
\end{figure}
\noindent {\small FIG. S2 (color online) $M$ vs. $B\parallel c$ at several $T$.  The arrows denote the direction of $B$-sweeps.}
\vspace{12pt}
%%%%%%%%%%%%%%%%%%%FIG-S2

Here we discuss in more detail the theoretical calculation of the magnetoresistance step size as a function of $T$
[Fig.~3(f)].  The calculation is based on the known mechanism of the negative MR in doped AF La$_2$CuO$_4$ [1].  By using a phenomenological model to account for the presence of skyrmions in the case of Li doping, we show that the $T$-dependence of the MR step size is reversed at low $T$, in agreement with the data [Fig.~3(f)].

It is widely accepted that the magnetic properties of AF undoped
La$_2$CuO$_4$, including the role of DM interaction [2], can
be fully accounted for by a semiclassical continuum field theory (the
quantum nonlinear sigma model) for the local spin moments on the Cu [3]. At very low doping, where the static and long-ranged AF order
is still dominant, a semiclassical treatment of the hole motion and of the
magnetic textures induced by the carriers is still justified. Within this
approach, it has been argued that the geometry and character of the dopants
can induce different local distortions of the AF background, in agreement
with the experimental observations.  For example, the formation of periodic
spin spirals [4], which is also supported via cluster Monte
Carlo method [5], can explain the incommensurate neutron
response [6] in a Sr-doped material very well.  On the other
hand, the substitution of Li$^+$ for Cu$^{2+}$ can, according to exact
diagonalization studies [7], stabilize a skyrmion, leading to the
precise signatures in magnetotransport discussed below. Finally, at higher
doping, when long-range magnetic order is completely suppressed and the
charge transport has a metallic character, various short-ranged spin and
charge modulations may emerge [8].

Here we focus on the case of very low doping.  To evaluate the MR in the
presence of skyrmion defects, we extend to the case of Li doping the
approach developed in Ref.~[1], which has been proved to
explain both qualitatively and quantitatively the MR data in Sr-doped AF
La$_2$CuO$_4$ [10 ,11]. As a starting point, we treat the
motion of the hole in the 2D AF background within a semiclassical approach
to the $t-t'-J$ model [5], where the hole spin is constrained to
be aligned to the AF background, so that the hole motion can be reduced to
the hopping of a spinless charge in a given sublattice,
with hole momentum near either of the two pockets $(\pi/2,\pm \pi/2)$.  The sublattice
degeneracy is further removed by orthorhombicity of the crystal, which
is large, and we assume that only the $b$ pocket $(\pi/2,-\pi/2)$ is
occupied [9].  At low
doping, the trapping Coulomb potential $V(r)$ provided by the dopants
(either Sr or Li) is unscreened, and the transport has an insulating
character.  The spin degrees of freedom are described using
the quantum nonlinear sigma model [3] in the
presence of the DM interaction [2], which successfully
describes the evolution of the magnetic order as a function of temperature $T$
and the applied magnetic field $B$.

The 2D picture for the hole motion is
partly modified by the presence of $B$ that tends to
align the weak FM moments and, as a consequence, induces free interlayer
tunneling $t_\perp(B)$ of the holes.
Indeed, as discussed in detail in Ref.\ [1], hopping
between planes is constrained by the direction of
ferromagnetic alignment of the spins.
At ${\bf B}=0$, the spins in neighboring layers
align antiferromagnetically within the $ac$-plane and
ferromagnetically within the $bc$-plane, while the opposite occurs
above the spin-flop transition, induced by the alignment of the WF moments along the
applied field. As a consequence, the hole dispersion is corrected as
$\delta\epsilon_{\bf k}\simeq - t_\perp \cos k_z \cos(k_x\pm k_y)/2$,
with the minus sign at $B<B_c$ and the plus sign at $B>B_c$. However,
since only the $b$ pocket is occupied, the hole momentum is near
$(\pi/2,-\pi/2)$, leading to a vanishing out-of-plane hopping for
$B<B_c$ and to a finite one for $B>B_c$.
The resulting Schr\"odinger equation
for the hole wave function can be written as a layered model [1]:
\begin{equation}
\label{model}
\left(-\frac{\nabla^2_r}{2 m^*}-V(r)\delta_{n0}\right)\psi^n-
t_\perp(B)[\psi^{n+1}+\psi^{n-1}-2\psi_n]=E\psi_n.
\end{equation}
Here $\psi^n$ is the envelope wave-function of the hole in the $n$-th
plane, and $m^*\approx 2m_e$ is the mass of low-energy quasiparticle
excitations in the AF background. For a shallow level the exact form of the
local potential is not important, so we can use a delta-like approximation
$V(r)=-g\delta(r)$ for the localizing potential $V(r)$ provided by the Li
dopant that acts only in the $n=0$ plane where the hole resides. For
$t_\perp=0$, the wave function of the hole with the full 2D kinetic
operator $\nabla^2_r$ is $\psi_0(r)\sim K_0(r/\xi_0)$, where $K_0$ is the
modified Bessel function of the second kind and $\xi_0$ is the localization
length, determined by $m^*$ and $g$. For any practical purpose, what
matters to the magnetotransport mechanism [1] is only the
asymptotic exponential decay of the hole wave function, $\psi_0(r) \sim
e^{-r/\xi_0}$ at $r\gtrsim \xi_0$. Thus, to allow for an analytical
treatment of the skyrmion case, we shall consider in what follows the
reduced one-dimensional problem associated with the radial part of the
envelope function. The strength $g$ of the localizing potential is adjusted
to reproduce for $t_\perp=0$ the localization length $\xi_0=15$ \AA \
extracted from the measured $T_0$.  In particular, $T_0$ was obtained from
the fit in Fig.~3(a) inset (a$_{1}$) and $\xi_0$ was
estimated from the expression $\xi_0=[13.8/(kT_{0}N_2)]^{1/2}$ valid for 2D
VRH [12], with the 2D density of states at the chemical potential
$N_{2}\approx (0.7-4.5)\times 10^{15}$(cm$^2$\, eV)$^{-1}$
($N=(2-13)$/(eV\, cell) [13]).

As we explained above, the $t_\perp$ term in Eq.~(\ref{model}) accounts for the interlayer hopping
process that becomes possible when the in-plane or out-of-plane magnetic
field is applied, causing a reorientation of the spins in neighboring
planes. In the perpendicular-field geometry and without skyrmions,
$t_\perp(B)=0$ for $B<B_c$,
while $t_\perp(B)=t_0$ for $B>B_c$, where $B_c$ is the critical field for
the spin-flop transition. We note that this description of
the hole motion is effective only in the presence of long-ranged AF
correlations, so that MR is expected to vanish at $T\gtrsim T_N$. While a
full treatment of the MR at high $T$
is beyond the scope of the present work, we partly accounted for this
effect by rescaling $t_\perp$ with the AF order parameter n$_0(T)$, which
has been calculated following Ref.~[2]. Note that this
does not affect the MR temperature dependence in the regime where skyrmions
form, because n$_0(T)$ here has a weak $T$ dependence, in agreement with
$\Delta M(T)\propto$ n$_0(T)$ [2] shown in
Fig.~3(f).

When skyrmions are not present, one recovers the case discussed in
Ref.~[1] for Sr-doped La$_2$CuO$_4$. Within a VRH scheme,
the decrease of the resistance across the spin-flop transition is evaluated
through an increase in the hole's probability $|\psi_B(r)|^2$ to propagate
at the VRH distance $r_h=(\xi_0/3)(T_0/T)^{1/3}$, hence $ \frac{R(B)-R(0)}{R(0)}=
 -(1-\frac{\mid\psi_{0}(r_{h})\mid^{2}}{\mid\psi_{B}(r_{h})\mid^{2}})$.
We note that even outside the regime where the
resistivity has strictly VRH character (\textit{i.e.} above 23~K in our
case) what matters for the $T$-dependence of the MR is just the
increase of the hopping probability $r_h(T)$ with decreasing $T$. Since we
are interested in discussing the role of skyrmions below $\sim 35$~K, we
will use for simplicity the VRH expression for $r_h(T)$ at all
temperatures.  This accounts qualitatively for the MR also at higher $T$, where a quantitative agreement with the data is not expected within the present approach, as explained above.

The formation of the skyrmion defect around the dopant site $r=0$ affects
this interlayer hopping process in a crucial way. Indeed, in the skyrmion
configuration the spins are reversed at the impurity site $r=0$.
Therefore, before the flop, spins in neighboring planes are ordered
ferromagnetically inside a distance of the order of the skyrmion core size
$\lambda$, while the opposite occurs above the spin-flop transition [see
Fig.\ 1(c)]. This can be modeled by means of a
distance-dependent tunneling process, \textit{i.e.} $t_\perp(r,B)=t_0
\theta(r-\lambda)$ at $B<B_c$, and $t_\perp(r,B)=t_0\theta(\lambda-r)$ at
$B>B_c$, where $\theta(x)=1$ for $x<0$ and zero otherwise, with $t_0\simeq
3.4$ meV [1].  Within this simplified, {\em hard} skyrmion model
for the interlayer hopping, according to which the spin quantization
axis is reversed discontinuously across the core size and the skyrmion
tail is removed, the parameter $\lambda$ turns out numerically much
smaller than the typical values obtained in Ref.~[7], where
$\lambda\approx 2-3$ in units of the lattice spacing. The above {\em hard}
skyrmion model could be refined, for example, by considering explicitly the
smoother $r$ dependence of a {\em soft} skyrmion in the spin-dependent
interlayer hopping, $t_\perp=(t_0/\sqrt{2}S)\sqrt{S^2+{\bf S}_{i}\cdot{\bf S}_{i+1}}$,
with the spin quantization axis for spins at the $i$-th layer, ${\bf S}_i$,
being reversed continuously across the core size. In this case, and as
considered in [7], one naturally expects larger numerical values
for $\lambda$. It is important to emphasize, however, that if one
includes also all magnetic anisotropies, the antisymmetric
Dzyaloshinskii-Moriya and XY interactions, as well as the interlayer
exchange $J_\perp$, the energy of the skyrmion far field will be increased
and the core size will be automatically reduced, as shown in [7]. In other
words, since the magnetic anisotropies enforce the $b$-orthorhombic
direction as the easy axis for the staggered moments, the core size is
automatically reduced in order to minimize the energy to be paid, by
the circulating hole, to generate a skyrmion distortion of the background,
thus justifying our choice for the {\em hard} skyrmion description. In any
case, since both models lead to analogous qualitative physics differing
only in the numerical value for $\lambda$, which in both cases is smaller
than the localization length $\xi_0$, we shall adopt, in what follows, the
{\em hard} skyrmion model, for the sake of simplicity.  We also note that
a small {\em hard}-skyrmion core size is expected from the $T$ dependence
of the magnetization step,
which is not affected by the skyrmion formation. Indeed, a significant
deformation of the spin background due to large skyrmions would decrease also the value of
the AF order parameter $n_0$, proportional to the magnetization step
shown in Fig.~3(f). At the same time, a point-like size of the magnetic
defects justifies also the neglect of the interactions between them, as
it is implicitly done here.

By solving again the model [Eq.~(\ref{model})],
one can see that the size of the MR step shrinks due to the presence of
skyrmions, which effectively reduce the interlayer hopping channel opened
by the spin reorientation. The size of this effect depends on the size
$\lambda$ of the skyrmion core, which we expect to increase with decreasing
$T$ to lower the energy of the quantum defect with respect to its
classical value [14].  In Fig.~3(f), we modeled the gradual increase
of $\lambda$ with decreasing $T$ as
$\lambda(T)=\lambda_0/(1+\exp((T-T_{sk})/T_w)$, where
$\lambda_0=0.08\xi_0$, $T_{sk}=17.5$~K and $T_w=15$~K. Here $T_{sk}$
represents the temperature scale where long-wavelength transverse
fluctuations of the AF order parameter (that are controlled by magnon gaps
around 10-20 K) disorder the skyrmion ground state.  The resulting $T$-dependence of the MR step size is in an excellent agreement with the data [Fig.~3(f)].
\vspace*{24pt}

%\begin{thebibliography}{43}
\small{
%\bibitem{kotov07}
\noindent [1] V. N. Kotov \textit{et al.},
%Sushkov, O. P., Silva Neto, M. B., Benfatto, L. \& Castro Neto, A. H.  Negative hopping magnetoresistance and
%dimensional crossover in lightly doped cuprate superconductors.
Phys. Rev. B \textbf{76}, 224512 (2007).

%\bibitem{benfatto06}
\noindent [2] L. Benfatto \textit{et al.},
%and Silva Neto, M. B.  Field dependence of the magnetic spectrum
%in anisotropic and Dzyaloshinskii-Moriya antiferromagnets. I. Theory.
Phys. Rev. B {\bf 74}, 024415 (2006).

%\bibitem{S1}
\noindent [3] S. Chakravarty \textit{et al.},
%Halperin, B.~I. \& Nelson D.~R.
%Two-dimensional quantum Heisenberg antiferromagnet at low temperatures.
Phys. Rev. B \textbf{39}, 2344 (1989).

%\bibitem{shraiman90}
\noindent [4] B. I. Shraiman \textit{et al.},
%and E. D. Siggia,
%Mobile vacancy in a quantum antiferromagnet: Effective Hamiltonian.
Phys. Rev. B \textbf{42}, 2485 (1990).

%\bibitem{sushkov}
\noindent [5] A. L\" uscher \textit{et al.},
%Misguich, G., Milstein, A. I. \& Sushkov, O. P.
%Local spin spirals in the N\' eel phase of La$_{2-x}$Sr$_{x}$CuO$_{4}$.
Phys. Rev. B {\bf 73}, 085122 (2006).

%\bibitem{matsuda}
\noindent [6] M. Matsuda \textit{et al.},
%Electronic phase separation in lightly doped La$_{2-x}$Sr$_{x}$CuO$_{4}$.
Phys. Rev. B {\bf 65}, 134515 (2002).

%\bibitem{haas96}
\noindent [7] S. Haas \textit{et al.},
%Zhang, F.-C., Mila, F. \& Rice, T.~M. Spin and charge texture around in-plane charge centers in the CuO$_2$ planes.
Phys. Rev. Lett. {\bf 77}, 3021 (1996).

%\bibitem{kivelson}
\noindent [8] S. A. Kivelson \textit{et al.},
%How to detect fluctuating stripes in the high-temperature superconductors.
Rev. Mod. Phys. \textbf{75}, 1201 (2003).
%I. P. Bindloss, E. Fradkin, V. Oganesyan, J. M. Tranquada,
%A. Kapitulnik, and C. Howald

\noindent [9] O. P. Sushkov \textit{et al.}, Phys. Rev. B \textbf{77}, 035124 (2008).

%\bibitem{andoMR03}
\noindent [10] Y. Ando \textit{et al.},
%Lavrov, A. N. \& Komiya, S.  Anisotropic magnetoresistance in lightly doped La$_{2-x}$Sr$_x$CuO$_4$: Impact of
%antiphase domain boundaries on the electron transport.
Phys. Rev. Lett. \textbf{90}, 247003 (2003).

%\bibitem{andoMR04}
\noindent [11] S. Ono \textit{et al.},
%Spin reorientation and in-plane magnetoresistance of lightly doped La$_{2-x}$Sr$_x$CuO$_4$ in magnetic
%fields up to 55 T.
Phys. Rev. B \textbf{70}, 184527 (2004).

%\bibitem{efros}
\noindent [12] B. I. Shklovskii and A. L. Efros, \textit{Electronic Properties of Doped
Semiconductors} (Springer-Verlag, Berlin, 1984).

%\bibitem{S2}
\noindent [13] R. L. Greene \textit{et al.},
%Maletta, H., Plaskett, T. S., Bednorz, J. G. \& M\"{u}ller, K. A. Evidence forelectron-electron correlations in La$_2$CuO$_4$ and La$_{2-x}$Sr$_{x}$CuO$_4$ superconductors.
Solid State Commun. \textbf{63}, 379 (1987).

%\bibitem{S3}
\noindent [14] J.~P. Rodriguez,
%Quantized topological point defects in two-dimensional antiferromagnets.
Phys. Rev. B {\textbf 39},  2906 (1995).
}
%\end{thebibliography}

\end{document}